\documentstyle[12pt,aaspp]{article}

\slugcomment{MIT-CSR-94-06, submitted to ApJ}

\begin{document}

\title{Lagrangian Evolution of the Weyl Tensor}

\author{Edmund Bertschinger}
\affil{Department of Physics, MIT 6-207, Cambridge, MA 02139}
   \and
\author{A. J. S. Hamilton}
\affil{Joint Institute for Laboratory Astrophysics; and
Department of Astrophysical, Planetary and Atmospheric Sciences;
Box 440, University of Colorado, Boulder, CO 80309}

\begin{abstract}
We derive the evolution equations for the electric and magnetic parts
of the Weyl tensor for cold dust from both general relativity and
Newtonian gravity.  In a locally inertial frame at rest in the fluid
frame, the Newtonian equations agree with those of general relativity.
We give explicit expressions for the electric and magnetic parts of the
Weyl tensor in the Newtonian limit.  In general, the magnetic part does
not vanish, implying that the Lagrangian evolution of the fluid is not
purely local.
\end{abstract}


\section{Introduction}

Gravity is a long-ranged force.  According to this Newtonian perspective,
the motion of a mass element is affected by distant elements.  In general
relativity, however, the motion of a freely-falling body is determined
by the local curvature of the spacetime manifold: in a locally inertial
frame, the motion of nearby mass elements is governed by the Riemann
curvature tensor.  The Newtonian and relativistic viewpoints are made
consistent by the fact that the Riemann tensor incorporates Newtonian
gravitational tidal fields.

In the Newtonian approach, tidal fields are obtained from the gradient
of the gravity vector, whose determination requires a sum over all mass
elements.  Thus, it came as a surprise to us that, in general relativity,
the tidal field evaluated at the position of a freely-falling mass element
might evolve according to a purely local equation --- provided that
certain stringent conditions are met.

Matarrese, Pantano, \& Saez (1993) and Croudace et al. (1994), following
earlier work of Barnes \& Rowlingson (1989), showed that if a quantity
known in general relativity as the magnetic part of the Weyl tensor
vanishes in the Newtonian limit, the Newtonian tidal field obeys a
local Lagrangian evolution equation until trajectories intersect.
In other words, the tidal field following a mass element changes with
time in a way depending only on local fluid variables --- the density
and velocity gradient.  Because these variables themselves obey
Lagrangian equations (local aside from the tidal field) --- it
seemed that the one might be able to evolve the fluid variables and
tidal field independently for all mass elements until trajectories
intersect.  In this situation, all nonlocal information is incorporated
into the initial value of the tidal field.

Bertschinger \& Jain (1994) used this fact to study the nonlinear evolution
of density perturbations in the expanding universe.  They noted (as did
Matarrese et al. 1994) that the magnetic part of the Weyl tensor did
not necessarily vanish in the Newtonian limit.  However, they assumed
that it could be approximately neglected, and solved the coupled nonlinear
fluid and tidal evolution equations for general irrotational motion
starting from the growing mode of cosmic density perturbations.  They
found that nonlinear coupling of the fluid shear and tidal field would
favor filamentary gravitational collapse as opposed to the sheetlike
pancake collapse predicted on the basis of kinematical theory
(Zel'dovich 1970).  This surprising conclusion rests on an important
unchecked assumption.

Until now there has been no Newtonian derivation of the tidal evolution
equation.  The equations of motion for the Riemann tensor in general
relativity have, instead, been projected into the local fluid frame.
This work was pioneered by Kundt \& Tr\"umper (1961), Hawking (1966),
and Ellis (1971, 1973).  It has resulted in a powerful covariant
Lagrangian fluid description of matter and gravitational fields.
This method has been applied to the evolution of cosmic density
fluctuations by Hawking (1966), Ellis \& Bruni (1989), Hwang \& Vishniac
(1990), and many later workers.  In linear perturbation theory the
magnetic part of the Weyl tensor vanishes for irrotational perturbations
so that the Lagrangian fluid equations reduce to local equations.
In cosmological perturbation theory the Lagrangian fluid approach,
while elegant and free of gauge ambiguities, offers no compelling
advantage compared with traditional Eulerian methods.

Traditionally, two approaches have been used to study gravitational
dynamics in cosmology.  The first is Eulerian: the fluid and
gravitational variables (mass density, velocity, etc.) are defined on
a grid and evolved according to partial differential equations.  This
method works well for a collisional fluid in which pressure forces
prevent the intersection of trajectories, but it breaks down for cold
dust (pressureless collisionless matter, e.g., cold dark matter) after
mass elements intersect.  The second method is that of Lagrangian
trajectories, exemplified by N-body simulations: the mass is discretized
into particles whose positions and velocities are integrated using the
appropriate equations of motion.

The Lagrangian fluid approach offers a third way:
the density and velocity gradients, but not necessarily the positions
and velocities, are integrated for individual mass elements.  Clearly,
this procedure is incomplete without computing the trajectories of
mass elements.  One can state the density of a given mass element
(provided that it is not superposed on another element) but, without
integrating the trajectories, one cannot say where it is.  While the
trajectories can be integrated here as in N-body simulations, this
presents an extra complication.  For some purposes, we would be happy
to know the density and velocity gradient of every mass element even
if their positions are unknown.

This explains the attraction of local Lagrangian evolution.  However,
the applicability of the method remains unclear as long as questions
remain about the magnetic part of the Weyl tensor.

In this paper we address three questions: Does the magnetic part of the
Weyl tensor vanish in the Newtonian limit? Can the Lagrangian evolution
equations for the Weyl tensor be derived in the Newtonian limit?  Are
they local in general?  In the following sections we show that the
answers are, respectively, not necessarily, yes, and no.

\section{Evolution of the Weyl tensor in General Relativity}

In this section we present the general relativistic derivation of
the evolution equations.  Although this material largely repeats the
work of Ellis (1971, 1973), it is a necessary prelude to the Newtonian
derivation.  Besides defining the Weyl tensor and other quantities,
we establish some notation that is used throughout this paper and
we clarify the conditions required to obtain the Newtonian limit of
general relativity.  We adopt the conventions and notations of
Misner, Thorne, \& Wheeler (1973), including metric signature +2.

The Weyl tensor is the traceless part of the Riemann curvature tensor
(Misner et al. 1973):
\begin{equation}
  C_{\mu\nu\kappa\lambda}\equiv R_{\mu\nu\kappa\lambda}-{1\over2}(
    g_{\mu\nu\kappa\sigma}R^\sigma_{\ \,\lambda}+
    g_{\mu\nu\sigma\lambda}R^\sigma_{\ \,\kappa})
    +{R\over6}\,g_{\mu\nu\kappa\lambda}\ ,
\label{weyl}
\end{equation}
where, for convenience in what follows, we have defined
\begin{equation}
  g_{\mu\nu\kappa\lambda}\equiv g_{\mu\kappa}g_{\nu\lambda}-
    g_{\mu\lambda}g_{\nu\kappa}\ .
\label{g4}
\end{equation}
The full Riemann tensor follows from the Weyl tensor and the Ricci
tensor $R_{\mu\nu}\equiv R^\kappa_{\ \,\mu\kappa\nu}$ and its trace
$R\equiv R^\mu_{\ \,\mu}$.  Through the Einstein field equations,
\begin{equation} R_{\mu\nu}-{R\over2}g_{\mu\nu}=8\pi G\,T_{\mu\nu}\ ,
\label{einstein}
\end{equation}
the Ricci tensor gives the contribution to the spacetime curvature
from local sources with energy-momentum tensor $T_{\mu\nu}$.  The
Weyl tensor gives the contribution due to nonlocal sources.  Therefore,
Newtonian tidal forces will be represented in the Weyl tensor.

In the most common formulation of general relativity, the Einstein
equations are regarded as field equations for the metric tensor
components, since the Ricci tensor can be obtained from the metric
tensor and its first and second derivatives.  Because the Riemann
tensor also follows from the metric and its derivatives, the Weyl tensor
can be calculated from the solution to the field equations.

However, we are interested in an alternative formulation of general
relativity in which the Weyl tensor is treated as the fundamental
geometrical quantity and the Ricci tensor follows algebraically from
the Einstein equations for a given distribution of energy and momentum.
This formulation requires equations of motion for the Weyl tensor
independently of the Einstein equations.  They follow from the
contracted Bianchi identities (Kundt \& Tr\"umper 1961).  In terms of
the Weyl and Ricci tensors, these are
\begin{equation}
  \nabla^\kappa C_{\mu\nu\kappa\lambda}=\nabla_{[\mu}R_{\nu]\lambda}
    +{1\over6}\,g_{\lambda[\mu}\nabla_{\nu]}R^\kappa_{\ \,\kappa}\ .
\label{bianchi}
\end{equation}
Gradient symbols denote the covariant derivative with respect to
$g_{\mu\nu}$.  Square brackets around a pair of indices denote
antisymmetrization, e.g., $A_{[\mu\nu]}\equiv\frac{1}{2}(A_{\mu\nu}
-A_{\nu\mu})$.  Symmetrized indices are surrounded by parentheses:
$A_{(\mu\nu)}\equiv\frac{1}{2}(A_{\mu\nu}+A_{\nu\mu})$.  Substituting
the Einstein equations (\ref{einstein}) into the Bianchi identities
(\ref{bianchi}) now provides field equations for the Weyl tensor:
\begin{equation}
  \nabla^\kappa C_{\mu\nu\kappa\lambda}=8\pi G
    \left(\nabla_{[\mu}T_{\nu] \lambda}+{1\over3}\,g_{\lambda[\mu}
    \nabla_{\nu]}T^\kappa_{\ \,\kappa} \right)\ .
\label{weyl-evol}
\end{equation}

Although the Weyl tensor has 256 components, only 10 of them are
independent in four dimensions.  It is convenient to incorporate
these 10 components into two symmetric second rank tensors using
the 4-velocity field $u^\mu(x)$ to split the Weyl tensor as follows:
\begin{equation}
  E_{\mu\nu}(u)\equiv u^\kappa u^\lambda C_{\mu\kappa\nu\lambda}\ ,\quad
  H_{\mu\nu}(u)\equiv {1\over2}\,u^\kappa u^\lambda\epsilon_{\alpha\beta
    \kappa(\mu}\,C^{\alpha\beta}_{\ \ \ \nu)\lambda}\ .
\label{weyl-split}
\end{equation}
These tensors, called the electric and magnetic parts of the Weyl tensor,
respectively, fully determine the Weyl tensor for any non-null normalized
$u^\mu$  (not just the 4-velocity of the matter).  Indeed, the full Weyl
tensor can be reconstructed from its electric and magnetic parts (Ellis
1971):
\begin{equation}
  C_{\mu\nu\kappa\lambda}=(g_{\mu\nu\alpha\beta}\,g_{\kappa\lambda
    \gamma\delta}-\epsilon_{\mu\nu\alpha\beta}\,\epsilon_{\kappa\lambda
    \gamma\delta})\,u^\alpha u^\gamma E^{\beta\delta}(u)
  +(\epsilon_{\mu\nu\alpha\beta}\,g_{\kappa\lambda\gamma\delta}+
    g_{\mu\nu\alpha\beta}\,\epsilon_{\kappa\lambda\gamma\delta})\,
    u^\alpha u^\gamma H^{\beta\delta}(u)\ .
\label{weyl-join}
\end{equation}
Equation (\ref{weyl-join}) is the exact inverse of equations
(\ref{weyl-split}) provided $g_{\mu\nu}u^\mu u^\nu=\pm1$ and the
same $u^\mu$ is used in both equations.  Ellis (1971) has a sign
error in the first term of his version of equation (\ref{weyl-join})
at the end of his section 4.3.3.  Note that we have used the fully
antisymmetric tensor $\epsilon_{\mu\nu\kappa\lambda}=(-g)^{1/2}\,
[\mu\nu\kappa\lambda]$, where $g$ is the determinant of $g_{\mu\nu}$
and $[\mu\nu\kappa\lambda]$ is the completely antisymmetric Levi-Civita
symbol defined by three conditions: (1) $[0123]=+1$, (2) $[\mu\nu\kappa
\lambda]$ changes sign if any two indices are exchanged, and (3)
$[\mu\nu\kappa\lambda]=0$ if any two indices are equal.  (Ellis
uses the tensor $\eta_{\mu\nu\kappa\lambda}=-\epsilon_{\mu\nu\kappa
\lambda}$.  We have compensated for the sign change in defining
$H_{\mu\nu}$.)

The Weyl tensor is completely independent of the 4-velocity $u^\mu$.
Only the electric and magnetic parts depend on $u^\mu$ --- in particular,
they are orthogonal to $u^\mu$: $E_{\mu\nu}u^\nu=H_{\mu\nu}u^\nu=0$.
Thus, in the local rest frame defined by $u^\mu$, only the spatial
components of $E_{\mu\nu}$ and $H_{\mu\nu}$ are non-vanishing.
Moreover, these tensors are trace-free: $E^\mu_{\ \,\mu}=
H^\mu_{\ \,\mu}=0$.  Each therefore has 5 independent components,
accounting for the 10 independent components of the Weyl tensor.

Equation (\ref{weyl-evol}) yield equations of motion for the electric
and magnetic parts of the Weyl tensor.  The results depend, of course,
on the energy-momentum tensor $T_{\mu\nu}$.  We assume that the matter
is a perfect fluid for which $T_{\mu\nu}=(\rho+p)u_\mu u_\nu+pg_{\mu\nu}$
with $u^\mu$ being the fluid 4-velocity and $\rho$ and $p$ being the
proper mass density and pressure, respectively, in the fluid frame.
Using the same 4-velocity to split the Weyl tensor into its electric
and magnetic parts, we obtain the following equations (Hawking 1966;
Ellis 1971):
\begin{eqnarray}
\label{div-E}
  \hbox{(div-$E$)}:\ \ &&
    P^\mu_{\ \,\alpha}P^\nu_{\ \,\beta}\nabla_\nu E^{\alpha\beta}
      +\epsilon^{\mu\nu\alpha\beta}u_\nu\sigma_{\alpha\gamma}H^\gamma_{\
      \,\beta}-3H^\mu_{\ \,\nu}\omega^\nu
      ={8\pi\over3}\,GP^{\mu\nu}\nabla_\nu\rho\ ,\\
\label{H-dot}
  (\dot H):\ \ &&
    P^\mu_{\ \,\alpha}P^\nu_{\ \,\beta}{DH^{\alpha\beta}\over d\lambda}
      -P^{\alpha(\mu}\epsilon^{\nu)\beta\gamma\delta}u_\beta\nabla_\gamma
      E_{\alpha\delta}-2u_\alpha a_\beta E_\gamma^{\ \,(\mu}\epsilon^
      {\nu)\alpha\beta\gamma}+\Theta H^{\mu\nu}\nonumber\\
    &&\quad\quad\quad\quad+P^{\mu\nu}(\sigma^{\alpha\beta}H_{\alpha\beta})
      -3H^{\alpha(\mu}\sigma^{\nu)}_{\ \ \alpha}
      +H^{\alpha(\mu}\omega^{\nu)}_{\ \ \alpha}=0\ ,\\
\label{div-H}
  \hbox{(div-$H$)}:\ \ &&
    P^\mu_{\ \,\alpha}P^\nu_{\ \,\beta}\nabla_\nu H^{\alpha\beta}
      -\epsilon^{\mu\nu\alpha\beta}u_\nu\sigma_{\alpha\gamma}E^\gamma_{\
      \,\beta}+3E^\mu_{\ \,\nu}\omega^\nu
      =-8\pi G(\rho+p)\omega^\mu\ ,\\
\label{E-dot}
  (\dot E):\ \ &&
    P^\mu_{\ \,\alpha}P^\nu_{\ \,\beta}{DE^{\alpha\beta}\over d\lambda}
      +P^{\alpha(\mu}\epsilon^{\nu)\beta\gamma\delta}u_\beta\nabla_\gamma
      H_{\alpha\delta}+2u_\alpha a_\beta H_\gamma^{\ \,(\mu}\epsilon^
      {\nu)\alpha\beta\gamma}+\Theta E^{\mu\nu}\nonumber\\
    &&\quad\quad\quad\quad+P^{\mu\nu}(\sigma^{\alpha
      \beta}E_{\alpha\beta})-3E^{\alpha(\mu}\sigma^{\nu)}_{\ \ \alpha}
      +E^{\alpha(\mu}\omega^{\nu)}_{\ \ \alpha}=
      -4\pi G(\rho+p)\sigma^{\mu\nu}\ .
\end{eqnarray}
We have adopted Ellis' names for these equations, inspired by their
similarity to the Maxwell equations (Ellis 1971, 1973).

In writing equations (\ref{div-E})--(\ref{E-dot}) we have defined
several new quantities.  The derivatives of $E^{\alpha\beta}$ and
$H^{\alpha\beta}$ are projected into the rest space of $u_\nu$ using
the projection tensor $P_{\mu\nu}\equiv g_{\mu\nu}+u_\mu u_\nu$.
The covariant derivative along the fluid worldline is $D/d\lambda\equiv
u^\nu\nabla_\nu$.  The velocity gradient has been decomposed into
the acceleration 4-vector $a_\nu$, expansion scalar $\Theta$, shear
tensor $\sigma_{\mu\nu}$, and vorticity tensor $\omega_{\mu\nu}$ or
its dual $\omega^\mu$:
\begin{eqnarray}
  &&\nabla_\mu u_\nu=-u_\mu{Du_\nu\over d\lambda}+
    P^\alpha_{\ \,\mu}P^\beta_{\ \,\nu}\nabla_\alpha u_\beta=
    -u_\mu a_\nu+{1\over3}\,\Theta P_{\mu\nu}+\sigma_{\mu\nu}+
    \omega_{\mu\nu}\,;\nonumber\\
  &&\Theta=\nabla_\mu u^\mu\ ,\ \
    \sigma_{\mu\nu}=\sigma_{(\mu\nu)}\ ,\ \
    \omega_{\mu\nu}=\omega_{[\mu\nu]}=
      \epsilon_{\mu\nu\alpha\beta}u^\alpha\omega^\beta\ .
\label{gradv}
\end{eqnarray}
In the fluid rest frame, $\omega_i$ is half the usual three-dimensional
vorticity $(\vec\nabla\times\vec v\,)_i$.  Ellis (1971) defines $\omega_
{\mu\nu}$ and $\omega^\mu$ with the opposite sign.

Equations (\ref{H-dot}) and (\ref{E-dot}) represent Lagrangian evolution
equations for the Weyl tensor: the time derivatives are taken along the
fluid worldlines.  Equations (\ref{div-E}) and (\ref{div-H}) are
constraint equations.  However, only the divergences of $E_{\mu\nu}$
and $H_{\mu\nu}$ are constrained; the ``curl'' parts (the gradient terms
in eqs. [\ref{H-dot}] and [\ref{E-dot}]) may be specified arbitrarily
on some initial hypersurface.  This is in contrast with Newtonian
gravity, where the gravitational field is fully determined from the
matter distribution through the static Poisson equation.  It is similar
to electromagnetism in that the gravitational field may contain a
source-free part corresponding (in the weak-field limit) to gravitational
radiation.

Equations (\ref{div-E})--(\ref{E-dot}) are fully covariant tensor
equations and can be applied in any coordinate system.  Because of the
key role played by the fluid 4-velocity, however, it is most convenient
to evaluate them in the fluid rest frame.  We shall assume that the
matter is cold (pressureless) dust, in which case the 4-acceleration
$a^\mu$ vanishes, so that the fluid rest frame is a locally inertial
frame.  Using locally flat coordinates in this frame, the equations of
motion for the electric and magnetic parts of the Weyl tensor become
\begin{equation}
  \nabla_j E^j_{\ \,i}-\epsilon_{ijk}\sigma^{jl}H^k_{\ \,l}
    -3H_{ij}\omega^j={8\pi\over3}\,G\nabla_i\rho\ ,
\label{divEij}
\end{equation}
\begin{equation}
  {dH_{ij}\over dt}+\nabla_k\,\epsilon^{kl}_{\ \ \,(i}E_{j)l}
    +\Theta H_{ij}+\delta_{ij}\,\sigma^{kl}H_{kl}
    -3\sigma^k_{\ \,(i}H_{j)k}-\omega^k_{\ \,(i}H_{j)k}=0\ ,
\label{Hijdot}
\end{equation}
\begin{equation}
  \nabla_j H^j_{\ \,i}+\epsilon_{ijk}\sigma^{jl}E^k_{\ \,l}
  +3E_{ij}\omega^j=-8\pi G\rho\,\omega_j\ ,
\label{divHij}
\end{equation}
\begin{equation}
  {dE_{ij}\over dt}-\nabla_k\,\epsilon^{kl}_{\ \ \,(i}H_{j)l}
    +\Theta E_{ij}+\delta_{ij}\,\sigma^{kl}E_{kl}
    -3\sigma^k_{\ \,(i}E_{j)k}-\omega^k_{\ \,(i}E_{j)k}=
    -4\pi G\rho\,\sigma_{ij}\ .
\label{Eijdot}
\end{equation}
Note that these equations assume $u^\mu=(1,\vec 0\,)$ at the point
where they are being applied because we are using locally flat coordinates
in the fluid rest frame, but the gradient of the 3-velocity does not
necessarily vanish.  In section 4 we shall furnish a Newtonian derivation
of these equations.

\section{Weyl Tensor for a Perturbed Robertson-Walker Spacetime}

As we saw in the preceding section, the electric and magnetic parts
of the Weyl tensor are only partly constrained by the matter distribution.
They are fixed, however, when the metric is specified.  In this section
we shall obtain the Weyl tensor components and its electric and magnetic
parts for a perturbed Robertson-Walker spacetime.

We start with the following line element:
\begin{equation}
    ds^2=a^2(\tau)\left\{-(1+2\psi)d\tau^2+2w_id\tau dx^i+\left[(1-2\phi)
    \gamma_{ij}+2h_{ij}\right]dx^idx^j\right\},\ \quad \gamma^{ij}h_{ij}=0\ .
\label{pmetric}
\end{equation}
The perturbations $\psi$, $\phi$, $w_i$, and $h_{ij}$ vanish for a
Robertson-Walker spacetime with expansion scale factor $a(\tau)$
($\tau$ is conformal time) and 3-metric $\gamma_{ij}$ (the metric
of a constant curvature space).  We treat the perturbations as being
small quantities.

As written, our metric is completely general.  To reduce the gauge
freedom we impose the following transversality constraints on $w_i$
and $h_{ij}$:
\begin{equation}
  \gamma^{ij}\nabla_iw_j=0\ ,\quad \gamma^{jk}\nabla_k h_{ij}=0\ ,
\label{pgauge}
\end{equation}
where $\nabla_i$ is the spatial covariant derivative relative to
$\gamma_{ij}$, whose inverse is $\gamma^{ij}$.  It can be shown that,
with these conditions, $\psi$ and $\phi$ are identical to the
gauge-invariant scalar mode variables $\Phi_A$ and $-\Phi_H$,
respectively, of Bardeen (1980).  Moreover, $w_i$ corresponds to the
``vector mode'' (gravitomagnetism) and $h_{ij}$ to the ``tensor mode''
(gravitational radiation).  As discussed in detail by Bertschinger
(1994), the gauge choice implied by equation (\ref{pgauge}) has several
advantages over other choices such as the synchronous gauge.  First,
it is essentially unique --- there is no residual gauge freedom
associated with spatially inhomogeneous redefinitions of the coordinates.
Second, by eliminating scalar mode contributions from $w_i$ and
$h_{ij}$, and vector mode contributions from $h_{ij}$, the physics of
the modes is simplified.  Finally, the perturbation variables are small
provided that physical curvature perturbations are small, so that the
coordinates $(\tau,x^i)$ are nearly identical to locally flat coordinates
scaled by a homogeneous expansion factor.  These last two advantages
facilitate relating calculations in this gauge to the Newtonian limit
(i.e., locally flat spacetime with slow source speeds relative to the
comoving frame).

Before giving the Weyl tensor components, we first present the equations
of motion for the metric perturbation variables obtained from the
Einstein equations.  These will provide intuition about the physics
of the different types of perturbations (scalar, vector, tensor) and
will facilitate comparison with the Newtonian limit.

The scalar and vector metric perturbation fields obey the following
equations (Bertschinger 1993) derived from the Einstein equations in
the limit of distance scales small compared with the curvature and Hubble
distances and with nonrelativistic shear stresses (the latter condition
implying $\psi=\phi)$:
\begin{equation}
  \vec\nabla\cdot\vec g=-4\pi Ga^2(\rho-\bar\rho)\ ,\quad
  \vec\nabla\times\vec g+\partial_\tau\vec H=0\ ,\quad
  \vec\nabla\cdot\vec H=0\ ,\quad
  \vec\nabla\times\vec H=-16\pi Ga^2\vec f_\perp\ ,
\label{gmaxwell}
\end{equation}
where
\begin{equation}
  \vec g\equiv-\vec\nabla\psi-\partial_\tau\vec w\ ,\quad
  \vec H\equiv\vec\nabla\times\vec w\ .
\label{ghfields}
\end{equation}
We use vector notation for three-dimensional vectors in the spatial
hypersurfaces with metric $\gamma_{ij}$.  Note that $\bar\rho(\tau)$
is the background density and $\vec f_\perp=(\rho\vec v\,)_\perp$
is the transverse energy current (transversality implying $\vec\nabla
\cdot\vec f_\perp=0$) evaluated in the comoving frame (so that $\vec v$
is the peculiar velocity, $\vec v=d\vec x/d\tau$ to lowest order in the
metric perturbations).

If one neglects $\vec w$, then equations (\ref{gmaxwell})--(\ref
{ghfields}) reduce to Newtonian gravity in comoving coordinates.
Because the source for the gravitomagnetic field $\vec H$ is
smaller by $O(v/c)$ than the source for Newtonian gravity, $\vec H$
usually is unimportant.  Note that equations (\ref{gmaxwell}) are
nearly identical with the Maxwell equations; the important
difference is the absence of longitudinal current and a displacement
current ($\partial_\tau\vec g\,$) term in the ``Amp\`ere'' law for
$\vec\nabla\times\vec H$.  This difference implies that both $\vec g$
and $\vec H$ are essentially static fields without radiation.

The absence of gravitational radiation from $\vec g$ and $\vec H$ is
what one expects for a spin-2 field (the graviton): radiation must
be present only in the spatial tensor mode.  Indeed, from the
Einstein equations one can show that the tensor mode obeys the wave
equation
\begin{equation}
  \left(\partial_\tau^2+2\eta\partial_\tau-\nabla^2+2K\right)h_{ij}=
    8\pi Ga^2\Sigma_{ij,\,\rm T}\ ,
\label{gwave}
\end{equation}
where $\eta\equiv d\log a/d\log\tau$, $\nabla^2=\gamma^{ij}\nabla_i
\nabla_j$ is the spatial Laplacian, $K$ is the spatial curvature
constant, and $\Sigma_{ij,\,\rm T}$ is the proper transverse-traceless
stress.

The field equations (\ref{gmaxwell})--(\ref{gwave}) show that for a
fluid with velocity $\vec v$, $w_i=\psi\times O(v/c)$ and $h_{ij}=\psi
\times O(v/c)^2$.  These relations will be important when we compare
with the Newtonian limit.

It is straightforward, though algebraically tedious, to compute the components
of the Weyl tensor for the metric given above.  The result is:
\begin{equation}
  C^0_{\ \,i0j}=-\frac{1}{2}\,D_{ij}(\psi+\phi)-\frac{1}{2}\,\dot W_{ij}
      +\frac{1}{2}(\partial_\tau^2+\nabla^2-2K)h_{ij}\ ,
\label{c0i0j}
\end{equation}
\begin{equation}
  C^0_{\ \,ijk}=(\nabla_k W_{ij}-\nabla_j W_{ik})+\frac{1}{4}(\nabla^2
      +2K)(\gamma_{ij}w_k-\gamma_{ik}w_j)+(\nabla_j\dot h_{ik}-\nabla_k
      \dot h_{ij})\ ,
\label{c0ijk}
\end{equation}
\begin{equation}
  C^i_{\ \,jkl}=\gamma^i_{\ \,jmn}\gamma^{pn}_{\ \ \,kl}\left[C^m_{\ \ 0p0}
      +(\nabla^2-3K)h^m_{\ \,p}\right]
      +\gamma^i_{\ \,jmn}\gamma^{pq}_{\ \ \,kl}\nabla_p\nabla^nh^m_{\ \,q}\ ,
\label{cijkl}
\end{equation}
where $\gamma^{ij}$ is used to raise the components of $\nabla_i$ and
$h_{ij}$ and we have defined several auxiliary quantities:
\begin{equation}
  D_{ij}\equiv\nabla_i\nabla_j-\frac{1}{3}\,\gamma_{ij}\nabla^2\ ,\quad
  W_{ij}\equiv\nabla_{(i}w_{j)}\ ,\quad
  \gamma^i_{\ \,jkl}\equiv \delta^i_{\ \,k}\gamma_{jl}-\delta^i_{\ \,l}
    \gamma_{jk}\ .
\label{auxil}
\end{equation}
All other components of the Weyl tensor follow from the symmetry relations
\begin{equation}
  C_{\mu\nu\kappa\lambda}=C_{[\mu\nu][\kappa\lambda]}=C_{\kappa\lambda
  \mu\nu}\ .
\label{symweyl}
\end{equation}
Note that because we are retaining only first-order terms in the metric
perturbations, the Weyl tensor components may be raised and lowered using
the unperturbed metric.  Thus, up to powers of the expansion factor,
we may regard equations (\ref{c0i0j})--(\ref{cijkl}) as giving us the
Weyl tensor components in locally flat coordinates.

The electric and magnetic parts of the Weyl tensor follow from equations
(\ref{weyl-split}).  We first give them using the 4-velocity of comoving
observers, $u^\mu=u_c^\mu=(a^{-1},\vec 0\,)$ for the metric of equation
(\ref{pmetric}).  Note that because the Weyl tensor is already first-order
in the metric perturbations, we need the 4-velocity only to zeroth-order.
We obtain (cf. Bruni, Dunsby, \& Ellis 1992, eqs. [113] and [115])
\begin{equation}
  E_{ij}(u_c)=\frac{1}{2}\,D_{ij}(\psi+\phi)+\frac{1}{2}\,\dot W_{ij}
      -\frac{1}{2}(\partial_\tau^2+\nabla^2-2K)h_{ij}\ ,
\label{Eij-c}
\end{equation}
\begin{equation}
  H_{ij}(u_c)=-\frac{1}{2}\,\nabla_{(i}H_{j)}+\nabla_k\epsilon^{kl}_
    {\ \ \,(i}\dot h_{j)l}\ .
\label{Hij-c}
\end{equation}
These components are given in the comoving coordinates of equation
(\ref{pmetric}).  To obtain the components in a locally flat coordinate
system at rest relative to the comoving frame (to first order in the
metric perturbations) one simply multiplies $E_{ij}$ and $H_{ij}$ by
$a^{-2}$.

In the Newtonian limit, $\psi=\phi$ is the Newtonian potential and
$E_{ij}$ is simply the gravitational tidal field (the traceless double
gradient of the potential, with a factor of $a^{-2}$ required to convert
the gradients from comoving to proper coordinates).  The magnetic part
of the Weyl tensor appears to have no Newtonian counterpart.  It depends
on gravitomagnetism and gravitational radiation, as do the corrections
to the electric part.

Next we must obtain the electric and magnetic parts of the Weyl tensor
in the fluid frame moving with 4-velocity $u^\mu$.  Two steps are
required for this computation.  First we must evaluate $E_{\mu\nu}$
and $H_{\mu\nu}$ using $u^\mu=a^{-1}(\gamma,\gamma\vec v\,)$ rather
than $u_c^\mu$ in equation (\ref{weyl-split}), where $\vec v$ is the
peculiar velocity and $\gamma\equiv(1-v^2)^{-1/2}$.  Then we must
transform from comoving coordinates to locally flat coordinates in the
fluid rest frame.  The result is
\begin{equation}
  E'_{ij}(u)=a^{-2}\Lambda^\mu_{\ \,i}(\vec v\,)\Lambda^\nu_{\ \,j}(\vec v\,)
    u^\kappa u^\lambda C_{\mu\kappa\nu\lambda}
\label{transf}
\end{equation}
and similarly for $H_{ij}$.  The Weyl tensor may be computed from
$E_{ij}(u_c)$ and $H_{ij}(u_c)$ using equation (\ref{weyl-join}).
A prime is used to indicate that the components are given in the
transformed frame moving with velocity $\vec v$ relative to the
unprimed (comoving) frame, and $\Lambda^k_{\ \,i}(\vec v\,)$ is the
Lorentz transformation corresponding to a boost $\vec v$: $\Lambda^0_
{\ \,0}=\gamma$, $\Lambda^0_{\ \,i}=\Lambda^i_{\ \,0}=\gamma v^i$,
$\Lambda^i_{\ \,j}=\delta^i_{\ \,j}+(\gamma-1)v^iv_j/v^2$.  We are
allowed to use special relativity here because we are working in
locally flat coordinates.  [The factor $a^{-2}$ converts $E_{\mu\nu}(u)$
from comoving to locally flat coordinates as described in the previous
paragraph.]

The two-stage transformation described above gives the following results:
\begin{equation}
  E'_{ij}=E_{ij}+2v_k\epsilon^{kl}_{\ \ \,(i}H_{j)l}+O(v/c)^2\ ,\quad
  H'_{ij}=H_{ij}-2v_k\epsilon^{kl}_{\ \ \,(i}E_{j)l}+O(v/c)^2\ ,
\label{transfEH}
\end{equation}
where, in each frame (primed and unprimed) the components are evaluated
using locally Minkowski coordinates.  (To reduce the clutter we have
dropped the arguments $u$ and $u_c$ from the primed and unprimed tensors,
respectively.)  Note that, although we began with a covariant definition
of $E_{\mu\nu}$ and $H_{\mu\nu}$, here we are evaluating them in two
different locally inertial frames such that in each frame the tensors
are purely spacelike.  Our results are reminiscent of the Lorentz
transformation of the ordinary electric and magnetic fields.  Ellis
(1973) shows how the electric and magnetic fields may be defined as
spacelike 4-vectors using the 4-velocity field to split the electromagnetic
field strength tensor in a way similar to what we have done here for the
Weyl tensor.

\section{Newtonian Evolution of the Weyl Tensor}

In the preceding sections we have derived the equations of motion
for the electric and magnetic parts of the Weyl tensor using general
relativity, and we have related these fields to the Newtonian gravitational
potential and other metric perturbation fields of a perturbed
Robertson-Walker spacetime.  In this section we provide a Newtonian
derivation of these results for cold dust.  We will work in a perturbed
Robertson-Walker spacetime, but a similar derivation can be carried
through in a non-cosmological setting.

We begin with the continuity and Poisson equations in comoving coordinates
(Bertschinger 1992):
\begin{equation}
  {\partial\delta\over\partial\tau}+\vec\nabla\cdot\left[(1+\delta)
    \vec v\,\right]=0\ ,\quad
  \nabla^2\phi=4\pi Ga^2(\rho-\bar\rho)\ .
\label{cont_pois}
\end{equation}
The mass density is written $\rho=\bar\rho(\tau)(1+\delta)$ and the
peculiar velocity is $\vec v=d\vec x/d\tau$.  Gradients are taken with
respect to comoving coordinates.  The cosmological Poisson equation
differs from its flat spacetime counterpart in that the mean background
density is subtracted from the source.  Equations (\ref{cont_pois}) are
valid to first order in $v/c$ on scales much smaller than the Hubble
distance $cH^{-1}$.  Note that these equations are valid even after orbits
intersect provided that $\delta$ and $\vec v$ are taken to be the total
density fluctuation and the average fluid velocity.

We define the Newtonian gravity vector in comoving coordinates,
\begin{equation}
  \vec g\equiv-\vec\nabla\phi\ ,\quad \vec\nabla\cdot\vec g=-4\pi Ga^2
    \bar\rho\,\delta\ .
\label{ggrav}
\end{equation}
This differs from equation (\ref{ghfields}) in that we neglect
$\partial_\tau\vec w$.  For slowly moving sources we expect
$\partial_\tau\sim\vec v\cdot\vec\nabla$ so that the neglected
contribution is $O(v/c)^2$ compared with the Newtonian gravitational
field.

For convenience we will decompose the velocity gradient into trace and
traceless symmetric and antisymmetric parts as in equation (\ref{gradv}):
$\nabla_iv_j=\frac{1}{3}\theta\delta_{ij}+\sigma_{ij}+\omega_{ij}$.
Note that we use $\theta=\vec\nabla\cdot\vec v$ for the divergence of
the peculiar velocity using comoving coordinates; $\Theta=\theta/a+3H$
is the proper divergence of the proper velocity including Hubble
expansion.  We decompose the gravity gradient in a similar manner:
\begin{equation}
  -\nabla_i\,g_j=\nabla_i\nabla_j\,\phi={1\over3}\,\delta_{ij}\,
    \nabla^2\phi+E_{ij}\ .
\label{gradg}
\end{equation}
This equation defines $E_{ij}$ in the Newtonian case; note that here,
as in equation (\ref{Eij-c}), we use the comoving components (Cartesian
so that $\gamma_{ij}$ becomes $\delta_{ij}$), which differ by a factor
$a^2(\tau)$ from the proper components.  Equations (\ref{Eij-c}) and
(\ref{gradg}) agree to $O(v/c)$.

Returning now to our derivation, we substitute $\vec g$ into the continuity
equation to obtain
\begin{equation}
  \vec\nabla\cdot\left(\partial\,a\vec g\over\partial\tau\right)=
    4\pi Ga^3\,\vec\nabla\cdot\vec f\ ,\quad
  \vec f\equiv \rho\,\vec v=\vec f_\parallel+\vec f_\perp\ .
\label{gcont}
\end{equation}
We have decomposed the mass current in the comoving frame into
longitudinal and transverse parts obeying $\vec\nabla\times\vec
f_\parallel=0$ and $\vec\nabla\cdot\vec f_\perp=0$.  Integrating
equation (\ref{gcont}) yields
\begin{equation}
  {\partial\vec g\over\partial\tau}+{\dot a\over a}\,\vec g=
    4\pi Ga^2\,\vec f_\parallel\ .
\label{gpart}
\end{equation}
The source involves only the longitudinal mass current because the
Newtonian gravity vector is longitudinal.  We rewrite equation
(\ref{gpart}) using the Lagrangian derivative following a fluid element,
$d/d\tau=\partial/\partial\tau+\vec v\cdot \vec\nabla$:
\begin{equation}
  {d\vec g\over d\tau}+{\dot a\over a}\,\vec g=
    \vec v\cdot\vec\nabla g+4\pi Ga^2\,\vec f_\parallel\ .
\label{gdot1}
\end{equation}

Before proceeding further we replace the transverse mass current
by a transverse vector field $\vec H$ obeying the field equations
\begin{equation}
  \vec\nabla\times\vec H=-16\pi Ga^2\,\vec f_\perp\ ,\quad
  \vec\nabla\cdot\vec H=0\ .
\label{Hfield}
\end{equation}
We recognize this as the gravitomagnetic field of equations
(\ref{gmaxwell}) and (\ref{ghfields}).  However, in the Newtonian
case we imply no relation between $\vec H$ and spacetime metric
perturbations; we simply regard $\vec H$ as a dynamical field
related to $\vec f_\perp$.

We further define a traceless tensor
\begin{equation}
  {\cal H}_{ij}\,\equiv -{1\over2}\,\nabla_j\,H_i
    +2v_k\epsilon^{kl}_{\ \ \,i}\nabla_j\,g_l
    =H_{ij}+\epsilon_{ijk}\,A^k\ ,
\label{defHijt}
\end{equation}
which we have decomposed into a symmetric part
\begin{equation}
  H_{ij}=-{1\over2}\,\nabla_{(i}H_{j)}-2\,v_k\,\epsilon^{kl}_
    {\ \ \,(i}E_{j)l}
\label{defHij}
\end{equation}
and an antisymmetric part with dual
\begin{equation}
  A_i={2\over3}\,v_i\,\nabla^2\phi-E_{ij}v^j+
    {1\over4}\left(\nabla\times\vec H\right)_i\ .
\label{defHi}
\end{equation}
To first order in $v/c$, equation (\ref{defHij}) agrees exactly with
the magnetic part of the Weyl tensor in the fluid frame (eqs.
[\ref{Hij-c}] and [\ref{transfEH}]).

For reference we provide some useful identities following from the
definitions of equations (\ref{ggrav})--(\ref{gradg}) and
(\ref{Hfield})--(\ref{defHi}).  First are two differential identities
for the divergence and curl of $\vec A$:
\begin{equation}
  \nabla_i A^i={2\over3}\,\theta\,\nabla^2\phi-\sigma^{ij}E_{ij}\ ,
\label{divAi}
\end{equation}
\begin{equation}
  \epsilon^{kl}_{\ \ \,i}\nabla_k A_l=-8\pi Ga^2\bar\rho\,\left(1+
    {1\over3}\,\delta\right)\,\omega_i-\epsilon_{ijk}\sigma^{jl}
    E^k_{\ \,l}+E_{ij}\omega^j\ .
\label{curlAi}
\end{equation}
Next are two differential identities for $H_{ij}$.  The first is
\begin{equation}
  \nabla_j H^j_{\ \,i}+\epsilon_{ijk}\sigma^{jl}E^k_{\ \,l}
  +3E_{ij}\omega^j=-8\pi Ga^2\rho\,\omega_j\ .
\label{divHij1}
\end{equation}
Aside from the factor $a^2$ present on the right-hand side because
we are using comoving rather than proper coordinates, this result
agrees {\it exactly} with equation (\ref{divHij}).  The last identity
will be useful in deriving equation (\ref{Eijdot}):
\begin{equation}
  \epsilon^{kl}_{\ \ \,i}\nabla_k H_{lj}=-\nabla_j A_i+{2\over3}\,
    \left(\nabla^2\phi\right)\,\nabla_jv_i+{2\over3}\,\theta\,E_{ij}
    +\delta_{ij}\,\sigma^{kl}E_{kl}-4\sigma^k_{\ \,(i}E_{j)k}-4
    \omega^k_{\ \,[i}E_{j]k}\ .
\label{curlHij}
\end{equation}

To derive equation (\ref{Eijdot}) we first substitute equation
(\ref{defHi}) into equation (\ref{Hfield}) and use equations (\ref{ggrav})
and (\ref{gdot1}) to get
\begin{equation}
  {d\vec g\over d\tau}+{\dot a\over a}\,\vec g=
    4\pi Ga^2\bar\rho\,\vec v+\vec A\ .
\label{gdot2}
\end{equation}
Taking the gradient gives
\begin{equation}
  {d\,\nabla_i\,g_j\over d\tau}+{\dot a\over a}\,\nabla_i\,g_j+
    (\nabla_i\,v^k)(\nabla_k\,g_j)=4\pi Ga^2\bar\rho\,\nabla_i\,v_j+
    \nabla_i\,A_j\ .
\label{gradgdot}
\end{equation}
The traceless symmetric part of this equation is
\begin{displaymath}
  {dE_{ij}\over d\tau}+{\dot a\over a}\,E_{ij}+{1\over3}\,\theta\,E_{ij}
    +{1\over3}\,\sigma_{ij}\,\nabla^2\phi+\sigma^k_{\ \,(i}E_{j)k}
    -{1\over3}\,\delta_{ij}\,\sigma^{kl}E_{kl}-\omega^k_{\ \,(i}E_{j)k}=
\end{displaymath}
\begin{equation}
  -4\pi Ga^2\bar\rho\,\sigma_{ij}-\nabla_{(i}A_{j)}+{1\over3}\,
    \delta_{ij}\,\nabla_kA^k\ .
\label{Eijdot2}
\end{equation}
The trace and traceless antisymmetric parts of equation (\ref{gradgdot})
give no further information because they simply reproduce the divergence
and curl of $\vec A$ that are given above.

Equation (\ref{Eijdot2}) involves the symmetrized gradient of the
antisymmetric part of ${\cal H}_{ij}$ defined by equation (\ref{defHijt}),
which can be replaced in favor of the antisymmetrized gradient of the
symmetric part $H_{ij}$ using equations (\ref{divAi}) and (\ref{curlHij}).
The necessary link is the identity
\begin{equation}
  \nabla_{(i}A_{j)}-{1\over 3}\,\delta_{ij}\,\nabla^k A_k=
    -\nabla_k\,\epsilon^{kl}_{\ \ \,(i}H_{j)l}+{2\over 3}\,\theta\,
    E_{ij}+{2\over3}\,\sigma_{ij}\,\nabla^2\phi-4\sigma^k_{\ \,(i}E_{j)k}
    +{4\over 3}\,\delta_{ij}\,\sigma^{kl}E_{kl}\ .
\label{gradA}
\end{equation}
When this is substituted into equation (\ref{Eijdot2}) we obtain
\begin{equation}
  {dE_{ij}\over d\tau}+{\dot a\over a}\,E_{ij}
    -\nabla_k\,\epsilon^{kl}_{\ \ \,(i}H_{j)l}+\theta E_{ij}
    +\delta_{ij}\,\sigma^{kl}E_{kl}-3\sigma^k_{\ \,(i}E_{j)k}
    -\omega^k_{\ \,(i}E_{j)k}
    =-4\pi Ga^2\rho\,\sigma_{ij}\ .
\label{Eijdot1}
\end{equation}
If the variables are converted to proper coordinates, $dt=ad\tau$,
$E_{ij}\to a^2E_{ij}$, etc., and one recalls that the comoving expansion
scalar is $\theta=a\Theta-3\dot a/a$, one obtains equation (\ref{Eijdot}).

We have succeeded in deriving the div-$H$ and $\dot E$ equations from
Newton's laws; what about the other two field equations for the Weyl
tensor?  Care is required because these equations involve terms
$O(v/c)^2$, as we can see by the following argument.  From their
definitions, we see that the units of $H_{ij}$ and $E_{ij}$ differ
by one power of a velocity: $H_{ij}\sim v E_{ij}$.  From their respective
field equations, one sees that $v$ must be of order the matter peculiar
velocity.  Dimensional analysis of equations (\ref{divEij}) and
(\ref{Hijdot}) shows that the $H_{ij}$ terms must be divided by $c^2$
and that they are $O(v/c)^2$ relative to the other terms.  (The same
analysis applied to eqs. [\ref{divHij}] and [\ref{Eijdot}] shows that
all terms are of the same order.)  In the strictly Newtonian limit
one would drop terms $O(v/c)^2$; for example, equations (\ref{cont_pois})
neglect relativistic corrections of this order.

Indeed, equations (\ref{divEij}) and (\ref{Hijdot}) are satisfied in
the Newtonian limit, as one can see by evaluating the divergence and
curl of $E_{ij}$ from equation (\ref{gradg}):
\begin{equation}
  \nabla_jE^j_{\ i}={2\over3}\,\nabla_i\nabla^2\phi\ ,\quad
  \nabla_k\,\epsilon^{kl}_{\ \ \,(i}E_{j)l}=0\ .
\label{E-diff}
\end{equation}
Using the Poisson equation for $\nabla^2\phi$, one obtains
\begin{equation}
  \nabla_j E^j_{\ \,i}={8\pi\over3}\,Ga^2\nabla_i\rho\ .
\label{divEij1}
\end{equation}
Aside from the use of comoving coordinates, this equation and the
second of equations (\ref{E-diff}) agree with equations (\ref{divEij})
and (\ref{Hijdot}) in the Newtonian limit.

This derivation is not fully satisfactory because we have not accounted
for all the terms in equations (\ref{divEij}) and (\ref{Hijdot}).
Fortunately, with a little care we can derive the full equations.

First, we must use the $E_{ij}$ and $H_{ij}$ obtained using general
relativity rather than Newtonian gravity, although we shall make the
simplifying assumptions that gravitational radiation and the
gravitational effect of shear stress, as well as terms that are
explicitly $O(v/c)^2$, can be neglected.  Using equations
(\ref{Eij-c}), (\ref{Hij-c}), and (\ref{transfEH}), in comoving
coordinates we obtain the results
\begin{equation}
  E_{ij}=\left(\nabla_i\nabla_j-\frac{1}{3}\,\delta_{ij}\,
    \nabla^2\right)\phi+\frac{1}{2}\,\dot W_{ij}
    +2v_k\epsilon^{kl}_{\ \ \,(i}H^{(0)}_{j)l}\ ,\quad
  H_{ij}=-\frac{1}{2}\,\nabla_{(i}H_{j)}
    -2v_k\epsilon^{kl}_{\ \ \,(i}E^{(0)}_{j)l}\ ,
\label{EHfields}
\end{equation}
where superscript $(0)$ indicates the quantity is to be evaluated setting
$v_k=0$.  We must include in equations (\ref{EHfields}) terms that are
explicitly first-order in $v_k$, even though we will evaluate the evolution
equations in the fluid rest frame for comparison with equations
(\ref{divEij}) and (\ref{Hijdot}), because we will need spatial derivatives
of $E_{ij}$ and $H_{ij}$.  This fact shows that we can, however, safely
ignore terms that are explicitly quadratic in $v_k$.

The divergence of $E_{ij}$ from equation (\ref{EHfields}) gives
\begin{equation}
  \nabla_j E^j_{\ \,i}={8\pi\over3}\,Ga^2\nabla_i\rho
    +\frac{1}{4}\nabla^2\dot w_i+\epsilon_{ijk}\sigma^{jl}H^k_{\ \,l}
    +3H_{ij}\omega^j\ ,
\label{divEij2}
\end{equation}
where we have retained only the terms that do not vanish when $v_k=0$.
If we can show that $\nabla^2\dot w_i=\partial_\tau(16\pi Ga^2f_{\perp
i})$ vanishes in the fluid frame, then equation (\ref{divEij2})
implies equation (\ref{divEij}).  We can show this using the following
trick.  In the fluid frame, $\vec f=\vec f_\parallel+\vec f_\perp=0$,
by definition.  Also, the fluid frame is freely-falling, so that the
fluid acceleration measured at the fluid element must vanish.  Equation
(\ref{gdot1}) then yields $\vec f_\parallel=0$, hence $\vec f_\perp=0$
and therefore $\nabla^2w_i=0$.  Since this result holds for all time
(the fluid continues to remain at rest in this inertial frame),
$\nabla^2\dot w_i=0$.  Equation (\ref{divEij}) then follows.

To derive the $\dot H$ equation we take the curl of $E_{ij}$ from
equation (\ref{EHfields}), again discarding terms that do not vanish
when $v_k=0$:
\begin{equation}
  \nabla_k\,\epsilon^{kl}_{\ \ \,(i}E_{j)l}=\frac{1}{2}\nabla_{(i}
    \dot H_{j)}-\theta H_{ij}-\delta_{ij}\,\sigma^{kl}H_{kl}
    +3\sigma^k_{\ \,(i}H_{j)k}+\omega^k_{\ \,(i}H_{j)k}\ .
\label{curlEij}
\end{equation}
In evaluating the time derivative term we must be careful to divide
the comoving components $H_j$ used here by $a$ to obtain the proper
components.  Then, using the fact that in the fluid frame $v_k=
dv_k/d\tau=0$, we obtain $dH_{ij}/d\tau=\partial_\tau H_{ij}
=-(\partial_\tau-\dot a/a)\frac{1}{2}\nabla_{(i}H_{j)}$.  Equation
(\ref{curlEij}) then reduces to
\begin{equation}
  {dH_{ij}\over d\tau}+{\dot a\over a}\,H_{ij}
    +\nabla_k\,\epsilon^{kl}_{\ \ \,(i}E_{j)l}
    +\theta H_{ij}+\delta_{ij}\,\sigma^{kl}H_{kl}
    -3\sigma^k_{\ \,(i}H_{j)k}-\omega^k_{\ \,(i}H_{j)k}=0\ .
\label{Hijdot1}
\end{equation}
Converting to proper coordinates and using the relation $\theta=
a\Theta-3\dot a/a$, we recover equation (\ref{Hijdot}).

It is interesting to note that in the Newtonian derivation of
the $H$-dot equation (14) --- but not the $E$-dot equation (16)
--- we had to assume vanishing acceleration in the fluid frame.
The reason is clear from the relativistic equations (9) and (11)
retaining the 4-acceleration.  By dimensional analysis, the
4-acceleration term in equation (11) must be divided by $c^2$
compared with the term in equation (9).  Thus, acceleration
effects are unimportant in the Newtonian tidal evolution equation
but are important in the evolution of $H_{ij}$.

\section{Conclusions}

We have shown the magnetic part of the Weyl tensor $H_{ij}$ does not
necessarily vanish in the Newtonian limit by deriving an explicit
expression for it --- equation (39).  As a consequence, the Lagrangian
evolution of the tidal field (the electric part of the Weyl tensor,
$E_{ij}$) is not purely local: it involves the gradient of $H_{ij}$
(eq. [\ref{Eijdot}]).  This term is of the same order in powers of
$v/c$ as the other terms affecting the tidal evolution.  No ambiguity
remains because we have obtained identical results using both general
relativity and Newtonian gravity.

It may be surprising that magnetic-like effects are significant for
the evolution of the tidal field.  After all, the Newtonian limit
neglects the gravitomagnetic force $m\vec v\times\vec H$ relative
to the gravitoelectric force $m\vec g$.  However, changes in the
tidal field are dependent on the motion of the fluid.  As in the
case of electromagnetism, electric fields in one frame transform into
magnetic fields in another frame.  Moreover, even if we choose a
frame so that $H_{ij}=0$ (not necessarily the fluid frame), this
condition is insufficient to restore locality because the evolution
of $E_{ij}$ depends on the gradient of $H_{ij}$.  Given the complexity
of the equations, at this time we cannot provide more physical insight
into our results.  However, it is clear that they are implied by mass
conservation and Newtonian gravity rather than some subtle relativistic
effects.

Although we have shown that the tidal evolution is, in general, nonlocal,
the Lagrangian fluid method still can be applied to study the nonlinear
evolution of self-gravitating mass.  One will simply have to evolve both
$E_{ij}$ and $H_{ij}$ and compute their gradients.  Since this requires
knowing the positions of mass elements, the trajectories will have to be
integrated simultaneously.  Although this greatly complicates matters,
there may be advantages in being able to compute the density and velocity
gradients of individual mass elements treated as a fluid rather than by
summing over particles weighted with a smoothing kernel.  After
trajectories intersect, however, the Lagrangian fluid method becomes
substantially more complicated.

Another possibility is that an alternative approximation may prove
useful in restoring locality.  Since we have obtained an explicit
Newtonian expression for $H_{ij}$, one can test alternative local
approximations for its gradient.  Of course, we still have not shown
how bad an approximation it is to neglect $H_{ij}$ altogether.
This can only be done by comparing with other approximations or
N-body simulations.  This work is currently underway.

Our results do not resolve the question of whether nonlinear gravitational
collapse favors the formation of prolate filaments as opposed to oblate
pancakes.  This question hinges critically on the behavior of $H_{ij}$
during gravitational collapse.  We leave this subject for future work.

\acknowledgments

We would like to thank John Bahcall for the hospitality of the
Institute for Advanced Study, where some of this work was performed.
We acknowledge useful discussions with Lam Hui, Bhuvnesh Jain,
Pawan Kumar, and Simon White.
This work was supported by NSF grant AST90-01762.

\end{document}